\documentclass[a4paper,12pt]{article}

\usepackage[latin1]{inputenc} 
\usepackage[T1]{fontenc}      
\usepackage[left=2.5cm,right=2.5cm,top=2.5cm,bottom=2.5cm,head=2cm]{geometry}        
\usepackage[english]{babel}  
                              
\usepackage{graphicx}
\usepackage{amsmath}

\usepackage[usenames,dvipsnames]{color}

\title{Around MOND}
\author{Anaëlle Hallé \\ ENS Ulm -University of St-Andrews}
\date{}

\begin{document}
\begin{titlepage}

\begin{center}

\large{Stage de recherche, FIP M1}

\vspace{3 cm}

\Huge{\bf{Autour de la MOND}} 

\vspace{0.5 cm}

\begin{figure}[h!]
\centering
\includegraphics[height=10cm]{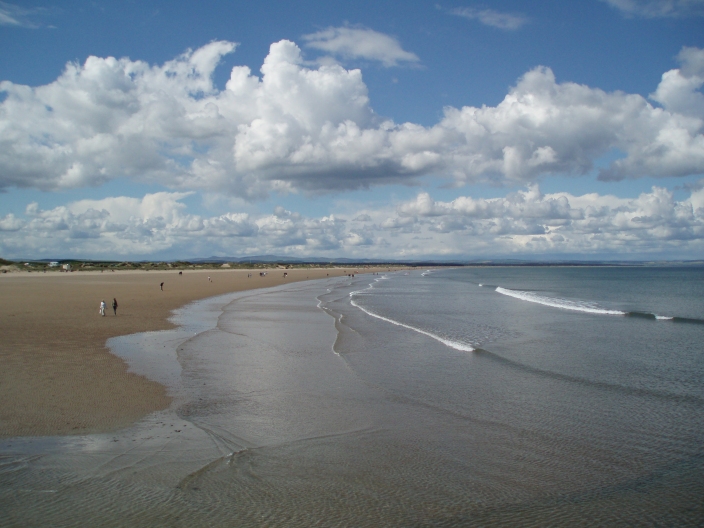} 
\end{figure}

\Huge{\bf{Around MOND}} 

\vfill

\large{Anaëlle Hallé \\ ENS Ulm}

\vfill

\large{Under the supervision of Dr HongSheng Zhao \\ University of St-Andrews, Scotland}

\vspace{1cm}

\large{February, 21$^{st}$ \ - \ August, 6 $^{th}$ 2007}

\end{center}
 
\end{titlepage}

\tableofcontents            


\clearpage

\pagestyle{plain}

~

\vfill

\begin{center}

\bf{Abstract}

\end{center}

The MOdified Newtonian Dynamics (MOND) is presented here, as well as a theory that can be linked to it: the theory of the Aether, a four-vector field breaking Lorentz invariance. The form of its Lagrangian is studied, then basic equations of the theory are rederived in a detailed way, and calculated for different metrics, exploring the impact of non-zero spatial terms of the Aether. A brief attempt of making the Aether Lagrangian depend on a scalar field is presented. An analytic solving of a galaxy model with an external field is described, which highlights the MONDian external field effect that breaks the strong equivalence principle.

\vspace{3 cm}

\begin{center}

\bf{Résumé}

\end{center}

On présente ici la gravitation Newtonienne modifiée (MOND) puis une théorie qui peut lui être reliée, celle de l'Éther, quadrivecteur brisant l'invariance de Lorentz. La forme du Lagrangien considérée pour l'Éther est étudiée, puis l'on redérive en détail les équations de base de la théorie, et on les calcule pour différentes métriques, en explorant l'impact de termes spatiaux non nuls pour l'Éther. On présente ensuite une tentative de faire dépendre d'un champ scalaire le Lagrangien de l'Éther. On s'intéresse également à une description analytique d'un modèle de galaxie sous un champ extérieur en gravitation Newtonienne modifiée, faisant ainsi ressortir l'effet de champ extérieur de la MOND, qui brise le principe d'équivalence fort.

\vfill

\newpage

\pagestyle{plain}

\section*{Introduction}

Gravitationnal physics is a great field of research and questionning for physicists. Newton's second law and General Relativity are no more sufficient today to account for astronomical observations without having to resort to uncertain forms of matter.  Astronomical observations show that the mass observed in the universe is far too small to account, for instance, for accelerations of objects in the outskirts of galaxies or for flat rotation curves of galaxies beginning at a certain radius.  A possibility to solve this problem is to resort to Dark Matter, named DM hereafter, which would consist of particles whose nature we can but speculate about today, except that they of course do not couple to photons. By adding DM, we can "artificially" raise gravitationnal fields, and by a clever fitting, adding to galaxies a halo of DM whose density decreases in $1/r²$, we can get flat rotation curves which are observed. \\

We can also think in a different way: maybe we are looking for something that does not exist. DM is indeed needed for observations to fit with our current theory of gravitation, but maybe this theory is not always correct. Newton's second law and General Relativity, that is build to reduce to it in a limit of low speed and weak slowly variating potentials, are successful in many cases. They describe very well our Solar System and phenomena in it such as the precession of the perihelion of Mercury or the Shapiro effect. But maybe it is not always true for all scales. \\

I worked during this internship under the supervision of Dr HongSheng Zhao working in the Physics and Astronomy department at the University of St-Andrews in Scotland. He's interested in DM and various aspects of the MOdified Newtonian Dynamics: galactic dynamics related issues, gravitationnal lensing, etc... He and his two Phd students Xufen Wu and Garry W. Angus form the MOND group at the University of St-Andrews. I had no predefined goal at the beginning: my supervisor wanted me first to get familiar with Aether theories poping up in the relativistic side of MOND and to rederive equations of the theory so as to carry out further work. As a member of this group, I read quite a lot about MOND and got in touch with some of the related work, and even worked on a specific MOND problem (Chapter 2). Meanwhile, I mostly focused on the Aether which is by itself linked to a huge part of current theoretical physics. \\

The next section is a short presentation of MOND, the first chapter is a study of Aether theories that can be linked to MOND and a report of my work about it, and the second chapter presents a "real MOND problem" linked to galactic dynamics and the "External Field Effect", which is a very important and controversial aspect of MOND, because it breaks the strong equivalence principle.
 
\section{MOND}

The MOdified Newtonian Dynamics emerged in the eighties. Milgrom wrote in 1983 several papers (see \cite{3}, \cite{4} and \cite{5}) to introduce it as an alternative to dark matter. The problems encountered seem to occur for low accelerations, so Milgrom proposed to modify the Newton's second law this way. 
\begin{equation}
m_g \mathbf{a} \mu \left( {\frac{a}{{a_0 }}} \right) = \mathbf{F}
\end{equation}

The value of $a_0$, about $1 \mathop A \limits^{ \ o} .s^{-2}$ was estimated by Milgrom in different ways, amongst which the fitting of rotation curves.

The function $\mu$ appearing here is the MOND interpolating function, which conveys the transition between the Newtonian and MONDian regimes.  The exact form of $\mu$ is not clear, different attempts have been made to fit observationnal data, but its asymptotic behaviour is:
\begin{itemize}
\item  $µ(x) = 1$ when $x \gg 1$ such that Newtonian gravity is recovered as expected when the gravitationnal field is far bigger than the acceleration scale $a_0$.
\item $µ(x) = x$ when $x \ll 1$ such that the formula will reduce to $\mathbf{g} g= a_0 \mathbf{g_N}$.
\end{itemize}

But it is not clear whether just gravity or inertia must be changed. We can choose to change just the gravitationnal field this way:
\begin{equation}
\mathbf{g} \mu \left( {\frac{g}{{a_0 }}} \right) = \mathbf{g_N} 
\end{equation} 

Is the whole physics in low acceleration modified? Some recent papers (see \cite{23} and \cite{24}) suggest that tests could be performed in the following years to try to check this on the Earth. This would anyway not be easy because it requires a high precision. It could rule out the MOND modification of inertia. \\
 
Such formulas are empirical. They describe lots of observations very well, but do not consist a theory. Bekenstein and Milgrom wrote in 1984 \cite{6} an action for non-relativistic MOND, so that by variating it, one can find a modified Newton-Poisson equation.
One can derive the Poisson equation by differenciation with respect to (w.r.t. hereafter) the gravitationnal potential of the following action:
\begin{equation}
S_N  = - \int {d^3 r\left[ {\rho \Phi _N  + \left( {8\pi G} \right)^{ - 1} \left( {\mathbf{\nabla}  \Phi _N } \right)^2 } \right]} \Rightarrow \mathbf{\nabla} ^2 \Phi _N  = 4\pi G\rho 
\end{equation}

whereas the following modified action:
\begin{equation}
S =  - \int {d^3 r[\rho \Phi  + \left( {8\pi G} \right)^{ - 1} a_0^2 F\left( {\frac{{\left( {\mathbf{\nabla}  \Phi } \right)^2 }}{{a_0^2 }}} \right) ] } 
\end{equation}

will generate the modified Poisson equation:
\begin{equation}
\mathbf{\nabla}  .( {\mu (  \frac{ |  \mathbf{\nabla}  \Phi  | }{a_0} )\mathbf{\nabla}  \Phi } )  =  4\pi G\rho, \quad  \mu (x) =F'(x^2)
\end{equation}

With this action, momentum, angular momentum and energy are conserved (see \cite{6}). The function $\mu$ and the constant $a_0$ are however still put \textit{ad hoc}, with no physical basis. 

\newpage

\pagestyle{headings}

\section{Modifications of the Einstein equations}

\subsection{Relativistic side of MOND}

A lot of work has been done to build a covariant theory around MOND. This theory would better be derivable from an action, namely to get conservation laws. Generally speaking, one builds a function of the dynamical variables of a theory, an action, that when variated w.r.t. these dynamicals variables gives the dynamic equations. Such a formalism is interesting because conservation laws follow then directly from symetry principles.\\

The Einstein-Hilbert action used for GR is:
\begin{equation}
S = \int {d^4 x\sqrt { - g} \left[ {\frac{R}{{16\pi G_N }} } \right]}  + S_M 
\end{equation}

where:
\begin{itemize}
\item  $g$ is the determinant of the metric $g_{\alpha \beta}$. This metric defines the geometry of spacetime. For two events of spacetime $P_1=(x_1^{\alpha})$ and $P_2=(x_2^{\alpha})$, the infinitesimal interval $ds$ between them is given by $ds^2=g_{\alpha \beta} dx^{\alpha} dx^{beta}$, where the $dx^{\alpha}$ are the coordinates difference between the events: $x_2^{\alpha}=(x_1^{\alpha})+dx^{\alpha}$.
The signature I took for all calculations in this report is $(-,+,+,+)$.
\item $R$ is the Ricci scalar. 
There are different conventions for the Rieman tensor $R^\lambda  \mathop {}\nolimits_{\mu \sigma \nu }$. I took for all the report:
\begin{eqnarray}
\left( {\nabla _\sigma  \nabla _\nu   - \nabla _\nu  \nabla _\sigma  } \right)V^\lambda   = R^\lambda  \mathop {}\nolimits_{\mu \sigma \nu } V^\mu  \\
R^\lambda  \mathop {}\nolimits_{\mu \sigma \nu }  = \partial _\sigma  \Gamma ^\lambda  _{\mu \nu }  - \partial _\nu  \Gamma ^\lambda  _{\mu \sigma }  + \Gamma ^\rho  _{\mu \nu } \Gamma ^\lambda  _{\rho \sigma }  - \Gamma ^\rho  _{\mu \sigma } \Gamma ^\lambda  _{\nu \rho } 
\end{eqnarray}
The Ricci tensor $R_{\mu \nu }$ and the Ricci scalar $R$ are::
\begin{equation}
R_{\mu \nu }  = R^\lambda  \mathop {}\nolimits_{\mu \lambda \nu }, \; R = g^{\mu \nu } R_{\mu \nu }
\end{equation}
These tensors describes the curvature of spacetime.
\footnote{The covariant derivatives and the Christoffel symbols $\Gamma ^\mu  _{\nu \rho }$ are defined the usual way: for a tensor $T^{\mu ...}_{\lambda ...}$,  $\nabla _{\nu}  T^{\mu ...}_{\lambda ...} = \partial _{\nu}  T^{\mu ...}_{\lambda ...} + \Gamma ^\mu  _{\nu \rho }  T^{\rho ...}_{\lambda ...} + \ ... \ - \Gamma ^\rho  _{\nu \lambda} T^{\mu ...}_{\rho ...} - \ ...$, with $\Gamma ^\mu  _{\nu \rho }  = g^{\mu \lambda } \left( {\partial _\nu  g_{\lambda \rho }  + \partial _\rho  g_{\lambda \nu }  - \partial _\lambda  g_{\nu \rho } } \right)$ } 
\item $S_M$ is the matter action that describes the matter distribution.
\end{itemize}

The light speed is $c=1$ in all the report. \\

Differenciation of this action with respect to the metric gives the Einstein equations:
\begin{equation}
\frac{{\delta S}}{{\delta g^{\alpha \beta } }} = 0 \Rightarrow G_{\alpha \beta }  =  8\pi GT_{\alpha \beta }^{matter} 
\end{equation}

with :
\begin{itemize}
\item $G_{\alpha \beta}=R_{\alpha \beta }  - \frac{1}{2}g_{\alpha \beta } R$: the Einstein tensor
\item $T_{\alpha \beta }^{matter}$: the stress-energy tensor of matter defined by: $\delta S_M  =  - \frac{1}{2}\int {d^4 x\sqrt { - g} T_{\mu \nu } \left( x \right)\delta g^{\mu \nu } \left( x \right)} $ This tensor describes the matter distribution.
\end{itemize}

These Einstein equations link the geometry of spacetime (by way of the curvature of its metric) to the matter distribution (by way of its stress energy tensor).\\

If we consider terms depending only on the curvature of space-time by the intermediate of the Ricci tensor as in GR, it can be shown that the Newtonian gravity will always be recovered \cite{7}, so we can't hope to recover MOND that way. We must add new degrees of freedom. After former attempts to build a fully covariant theory which encountered problems such as violation of causality, in 2004, Bekenstein proposed a theory named TeVeS, for Tensor Vector Scalar (see \cite{7}) in which he considers three distinct degrees of freedom: a tensor (the metric), a vector and a scalar field.  Zlosnik et al. noticed in 2005 that this theory could be reduced to a purely vector-tensor theory (see \cite{13}). \\

I did not work on TeVeS at a whole, but just on the Aether, the four-vector field that appears in it. The use of such a vector is embedded in a series of attempts to build Vector-Metric theories of gravity. Such a vector was considered by Will and Nordvedt in 1972 \cite{21}, amongst other means of exploring the possibilty and the impact of "preferred frames" in gravitationnal theories. Once again, this upsets physics because it breaks Lorentz invariance, which reads that there is no preferred frame. Lorentz invariance can't be tested uniformly because a parameter of the Lorentz group is unbounded, so we don't know if it holds at each scale. What's more, this invariance leads to divergences in quantum fields theories, so exploring breakings of it is "allowed" and tempting. \\

Breaking the invariance globally by choosing for instance a universal preferred rest frame or fixed background tensors is not satisfying because one can't this way preserve general covariance that is required for the Einstein equations to hold. One should give up GR or any modification of it, which is not theoretically appealing (see \cite{8}). One therefore tries to break this invariance locally. A four vector-field with a non-vanishing time component is one of the most simple toys that can be used to do so. It will select a preferred direction, a local dynamical preferred rest frame at each point of space-time, and can therefore be seen like a four-velocity of a fluid present evereywhere (that's why it is called the Aether). Such vectors coupling to matter were ruled out by experiments, but one can choose to consider a vector that couples only to the metric. A lot of work about such vectors has been done in recent years, especially by  Kostelecky (\cite{22}), Jacobson, Mattingly and Elling (\cite{8}, \cite{9}, and \cite{15} ), Lim and Carroll (\cite{10} \cite{11}), and Ferreira and Zlosnik (\cite{12} \cite{13}).

\subsection{Aether Lagrangian}

Understanding why certain people use a given form of the Lagrangian, and why other use something else, include other terms or not requires to look closely at this Lagrangian. What follows in this section is an explanation of the logics of the Lagrangians used, after a bibliographic research amongst various papers, especially by Will and Norvedt (see \cite{16} \cite{21}), Jacobson, Mattingly and Elling (see \cite{8}, \cite{9}, and \cite{15}) .\\

To write down an action for a vector-metric theory, one can look for a Lagrangian scalar density (scalar so that it is independant of the volume of integration in the action) which will, for considerations of simplicity, give a linear equation for the vector field and second order equations at most. The most general Lagrangian scalar density is thus:
\begin{eqnarray}
L \left( A,g \right) = a_0 + a_1 R + a_2 A^{\alpha} A_{\alpha} + a_3 A^{\alpha} A_{\alpha} R + a_4 A^{\alpha} A^{\beta} R_{\alpha \beta} +a_5 ( \nabla _{\alpha} A_{\beta} ) ( \nabla ^{\alpha} A^{\beta} )   \nonumber \\
+ a_6 \left(\nabla _{\alpha} A^{\alpha} \right)^2 + a_7  ( \nabla _{\alpha} A_{\beta} ) ( \nabla ^{\beta} A^{\alpha} ) +a_8 ( A^{\beta}\nabla _{\beta} A^{\alpha} ) ( A^{\gamma}\nabla _{\gamma} A_{\alpha} )
\end{eqnarray}

the $a_i$ being constants. All the possible combinations of indices for the quadratic terms in covariant derivatives are present here. One can notice that the Aether part of the Lagrangian density involves covariant derivatives and thus Christoffel symbols containing first derivatives of the metric, so this part of the action will contribute to the metric kinetic terms too. \\

But one can simplify this Lagrangian density. First, the difference of the term with the coefficient $a_6$ and the one with $a_7$ with the one in $a_4$ is $\nabla _{\alpha} \left( A^{\alpha} \nabla ^{\beta} A^{\beta} - A^{\beta} \nabla ^{\beta} A^{\alpha} \right)$, ie a total divergence, which will thus contribute just by a boundary term, according to the Stokes theorem. We can therefore choose not to consider the term in $a_4$.\\ 

The exact form of the Aether field depends on what one is looking for. If one wants not to keep Lorentz invariance at all or do not mind to do so, one can choose to fix the norm of the vector so that it will always have a non-vanishing timelike component and will therefore always be Lorentz-violating\footnote{The Aether will thus carry a non-linear representation of the local Lorentz group because it will not belong to a vector-space because its norm being fixed, it will take its value on an hyperboloïd of the tangent space of an event of space-time.}. The additionnal constraint can be enforced in a maybe non-appealing way,  ie using a non-dynamic Lagrange multiplier $\lambda$, but will also simplify the equations a lot.\\

If one fixes the norm, the terms in $a_2$ and $a_3$ play thus the same role as the one respectively in $a_0$ and $a_1$ and are thus useless.\\

The $a_0$ term plays simply the role of a cosmological constant (Namely we get a term proportionnal to $G_{\alpha \beta} + \Lambda g_{\alpha \beta}$ by differenciation of the action $S = \int {d^4 x\sqrt { - g} (a_0 + a_1 R)}$ if $\Lambda= - \frac{a_0}{2 a_1}$.).\\   

We are left thus left with:
\begin{eqnarray}
L \left( A,g \right) = a_1 R + a_5 ( \nabla _{\alpha} A_{\beta} ) ( \nabla ^{\alpha} A^{\beta} ) + a_6 \left(\nabla _{\alpha} A^{\alpha} \right)^2 + a_7  ( \nabla _{\alpha} A_{\beta} ) ( \nabla ^{\beta} A^{\alpha} ) \nonumber \\
+a_8 ( A^{\beta}\nabla _{\beta} A^{\alpha} ) ( A^{\gamma}\nabla _{\gamma} A_{\alpha} )
\end{eqnarray}

One can so consider the action:
\begin{eqnarray}
S = \frac{1}{{16\pi G_N }}\int {d^4 x\sqrt { - g} \left[ {R + K^{\alpha \beta } \mathop {}\nolimits_{\gamma \sigma } \nabla _\alpha  A^\gamma  \nabla _\beta  A^\sigma   + \lambda \left( {A^\alpha  A_\alpha   + 1} \right)} \right]}  \\ 
K^{\alpha \beta } \mathop {}\nolimits_{\gamma \sigma } = c_1 g^{\alpha \beta } g_{\gamma \sigma }  + c_2 \delta _\gamma ^\alpha  \delta _\sigma ^\beta   + c_3 \delta _\sigma ^\alpha  \delta _\gamma ^\beta   + c_4 A^\alpha  A^\beta  g_{\gamma \sigma } 
\end{eqnarray}

This action is the one which was considered by Jacobson, Eling and Mattingly. Notice that dropping the terms in $c_2$ and $c_4$ and considering $c_3=-c_1$, we find $K^{\alpha \beta } \mathop {}\nolimits_{\gamma \sigma } \nabla _\alpha  A^\gamma  \nabla _\beta  A^\sigma= \frac{c_1}{2} F_{\alpha \sigma} F^{\alpha \sigma}$, where $F_{\alpha \sigma}$ is the antisymmetric Maxwell tensor defined by $F_{\alpha \sigma}=\nabla _\alpha  A_\sigma - \nabla _\sigma  A_\alpha$. This simplification was used by Jacobson and by Bekenstein in TeVeS.\\

I focused on an action involving a general function $F$ of the Aether kinetic terms. 
\begin{eqnarray}
S = \frac{1}{{16\pi G_N }}\int {d^4 x\sqrt { - g} \left[ {R + M^2 \mathcal{F \left( \mathcal{K} \right) } + \lambda \left( {A^\alpha  A_\alpha   + 1} \right)} \right] }  \\
\mathcal{K} = \frac{1}{M^2} K^{\alpha \beta } \mathop {}\nolimits_{\gamma \sigma } \nabla _\alpha  A^\gamma  \nabla _\beta  A^\sigma \\
K^{\alpha \beta } \mathop {}\nolimits_{\gamma \sigma } = c_1 g^{\alpha \beta } g_{\gamma \sigma }  + c_2 \delta _\gamma ^\alpha  \delta _\sigma ^\beta   + c_3 \delta _\sigma ^\alpha  \delta _\gamma ^\beta   + c_4 A^\alpha  A^\beta  g_{\gamma \sigma } 
\end{eqnarray}

This action was considered recently by Zlosnik et al. (see \cite{12}), but with no term in $c_4$ to simplify. 

\subsection{Fields equations}

I rederived the equations from Zlosnik et al. Aether paper \cite{12}. For I had not studied the Lagrangian version of GR before, I had to get used to this formalism, especially to various variations of the action such as w.r.t. the metric and the subtulties of calculations. \\

What must be borne in mind when carrying out the variations is that the two dynamical degrees of freedom considered are the inverse metric $ g^{\mu \nu }$ and the contravariant Aether vector field $ A^{\mu} $. The contravariant Aether is chosen (and not the covariant one) just because once one has chosen to variate the action w.r.t. $g^{\mu \nu }$, the result of this variation will be simpler seeing the form of $K^{\alpha \beta } \mathop {}\nolimits_{\gamma \sigma }$ because we have :
\begin{equation}
\frac{{\delta A^\mu  }}{{\delta g^{\alpha \beta } }} = 0
\end{equation}

whereas:
\begin{eqnarray}
g_{\mu \rho } g^{\rho \sigma }  = \delta _\mu ^\sigma  \Rightarrow \delta g_{\mu \nu }  =  - g_{\mu \rho } g_{\nu \sigma } \delta g^{\rho \sigma } \\
and \; so \; \frac{{\delta A_\mu  }}{{\delta g^{\alpha \beta } }} = A^\nu  \frac{{\delta g_{\mu \nu } }}{{\delta g^{\alpha \beta } }} =  - g_{\mu \alpha } A_\beta  
\end{eqnarray}

The vector equation is obtained by varying the action w.r.t. $A^{\mu}$:
\begin{equation}
\frac{{\delta S}}{{\delta A^\alpha  }} = 0 \Rightarrow \nabla _\alpha  (F'J^\alpha  \mathop {}\nolimits_\beta  ) - F'y_\beta   = 2\lambda A_\beta  
\end{equation}

where:
\begin{itemize}
\item $F' = \frac{{dF}}{{dK}}$  
\item $J^\alpha  \mathop {}\nolimits_\sigma$ is a current: $J^\alpha  \mathop {}\nolimits_\sigma   = (K^{\alpha \beta } \mathop {}\nolimits_{\sigma \gamma }  + K^{\beta \alpha } \mathop {}\nolimits_{\gamma \sigma } )\nabla _\beta  A^\gamma$
(One can notice that if $K^{\alpha \beta } \mathop {}\nolimits_{\sigma \gamma } =  K^{\beta \alpha } \mathop {}\nolimits_{\gamma \sigma } $ like here, for the case considered, $J^\alpha  \mathop {}\nolimits_\sigma=2  K^{\alpha \beta } \mathop {}\nolimits_{\sigma \gamma } \nabla _\beta  A^\gamma$, but defining the current this way preserves the generality of the equations for a tensor which does not have such a symmetry.)
\item $y_\beta   = \nabla _\sigma  A^\eta  \nabla _\gamma  A^\xi  \frac{{\delta (K^{\sigma \gamma } \mathop {}\nolimits_{\eta \xi } )}}{{\delta A^\beta  }}$ 
\end{itemize}

\medskip 

We can get the Lagrange multiplier $\lambda$ from here.\\

Variating the action w.r.t. $\lambda$ will give the constraint on the norm:
\begin{equation}
A^{\alpha}A_{\alpha} = -1
\end{equation}

For the variation of the action $S = \int {d^4 x\sqrt { - g} L} $ w.r.t. the contravariant metric, one must notice that $\frac{{\delta S}}{{\delta g^{\alpha \beta } }} = \int {d^4 x} \sqrt { - g} \left( {\frac{{\delta L}}{{\delta g^{\alpha \beta } }} - \frac{1}{2}g_{\alpha \beta } L} \right)$ where one uses the fact that: $\delta g = gg^{\mu \nu } \delta g_{\mu \nu }  =  - gg_{\mu \nu } \delta g^{\mu \nu }$, $g$ being the determinant of the contravariant metric.\\

The symmetry of $K^{\alpha \beta } \mathop {}\nolimits_{\sigma \gamma }$ simplifies the equations:
\begin{eqnarray}
 \frac{{\delta \left( {M^2 F\left( K \right)} \right)}}{{\delta g^{\alpha \beta } }} &=& F' [ {Y_{\alpha \beta }  + K^{\sigma \gamma } \mathop {}\nolimits_{\eta \xi } \frac{{\delta \left( {\nabla _\sigma  A^\eta  } \right)}}{{\delta g^{\alpha \beta } }}\nabla _\gamma  A^\xi   + K^{\sigma \gamma } \mathop {}\nolimits_{\eta \xi } \nabla _\sigma  A^\eta  \frac{{\delta \left( {\nabla _\gamma  A^\xi  } \right)}}{{\delta g^{\alpha \beta } }}} ]  \nonumber \\ 
  &=& F' [ {Y_{\alpha \beta }  + J^\sigma  \mathop {}\nolimits_\eta  \frac{{\delta \left( {\nabla _\sigma  A^\eta  } \right)}}{{\delta g^{\alpha \beta } }}} ] 
\end{eqnarray}

with $Y_{\alpha \beta }$ a functionnal derivative defined by:
\begin{equation}
Y_{\alpha \beta }  = \nabla _\sigma  A^\eta  \nabla _\gamma  A^\xi  \frac{{\delta (K^{\sigma \gamma } \mathop {}\nolimits_{\eta \xi } )}}{{\delta g^{\alpha \beta} }}
\end{equation}

The variation of the covariant derivative of the contravariant Aether requires to variate the Christoffel symbol (only): 
\begin{equation}
\frac{{\delta \left( {\nabla _\sigma  A^\eta  } \right)}}{{\delta g^{\alpha \beta } }} = \frac{{\delta \left( {\partial _\sigma  A^\eta   + \Gamma _{\sigma \rho }^\eta  A^\rho  } \right)}}{{\delta g^{\alpha \beta } }} = \frac{{\delta \left( {\Gamma _{\sigma \rho }^\eta  } \right)}}{{\delta g^{\alpha \beta } }}A^\rho  
\end{equation}

And we have $\delta \left( {\Gamma _{\sigma \rho }^\eta  } \right) = \frac{{g^{\eta \tau } }}{2}\left( {\nabla _\sigma  \delta g_{\rho \tau }  + \nabla _\rho  \delta g_{\sigma \tau }  - \nabla _\tau  \delta g_{\sigma \rho } } \right)$ (see Weinberg, \cite{1}) so one eventually find:
\begin{equation}
F'J^\sigma  \mathop {}\nolimits_\eta  \frac{{\delta \left( {\nabla _\sigma  A^\eta  } \right)}}{{\delta g^{\alpha \beta } }} =  - \frac{1}{2} \nabla _\sigma  (\mathcal{F'}(J_{(\alpha } \mathop {}\nolimits_{}^\sigma  A_{\beta )}  - J^\sigma  {}_{(\alpha }A_{\beta )}  - J_{(\alpha \beta )} A^\sigma  ))
\end{equation}

dropping divergence terms which would once more contribute only by boundary terms. The brackets denote symmetrization, ie for instance $J_{\left( {\alpha \beta } \right)}  = \frac{1}{2}\left( {J_{\alpha \beta }  + J_{\beta \alpha } } \right)$.\\

We have also:
\begin{equation}
\frac{{\delta A^\mu  A_\mu  }}{{\delta g^{\alpha \beta } }} =  - A_\alpha  A_\beta  
\end{equation}

So putting all of that together, one eventually find:
\begin{equation}
G_{\alpha \beta }  =  8\pi GT_{\alpha \beta }^{matter} + T_{\alpha \beta }^{Aether} 
\end{equation}

where $T_{\alpha \beta} ^{Aether}$ is the Aether stress-energy tensor:
\begin{eqnarray}
T_{\alpha \beta} ^{Aether}  = \frac{1}{2}\nabla _\sigma  (\mathcal{F'}(J_{(\alpha } \mathop {}\nolimits_{}^\sigma  A_{\beta )}  - J^\sigma  {}_{(\alpha }A_{\beta )}  - J_{(\alpha \beta )} A^\sigma  ))  \nonumber \\
- \mathcal{F'}Y_{(\alpha \beta )} + \frac{1}{2}g_{\alpha \beta } M^2 \mathcal{F} + \lambda A_\alpha  A_\beta  
\end{eqnarray}

\newpage

\subsection{Exploring different regimes}

Once these equations of motions are obtained, one can calculate them for different metrics to explore different regimes.\\

My supervisor made me consider a Friedan-Robertson-Walker (FRW hereafter) pertubed metric such that: 
\begin{equation}
ds^2  =  - (1 + 2 \epsilon \phi )dt^2  + a(t)^2 (1 + 2 \epsilon \psi )(dx^2  + dy^2  + dz^2 )
\end{equation}
This metric is a perturbed form of the one of a homogeneous and spatially isotropic universe (also spatially flat here, ie with no curvature parameter). $\phi$ and $\psi$ are scalar gravitational potentials (that are identified in the non-relativistic limit), and $a(t)$ is the cosmic scale factor (it can be described as setting the scale of the geometry of space).\\

We can recover:
\begin{itemize}
\item the non relativistic limit by neglecting time derivatives and taking $a(t)=1$ to have $ds^2  =  - (1 + 2 \epsilon \phi )dt^2  + (1 + 2 \epsilon \psi )(dx^2  + dy^2  + dz^2 )$ 
\item a homogeneous and isotropic universe by taking $\epsilon=0$ to have $ds^2  =  - dt^2  + a(t)^2 (dx^2  + dy^2  + dz^2 )$ 
\end{itemize}

In the following, the equations are developped up to orders in $\epsilon$, but $\epsilon$ is not kept for a better lightness. Some equations that are not enlightning to follow the report are put in Annex A.

\paragraph{Einstein tensor} 

Up to linear order in $\epsilon$ we find:

\begin{itemize}
\item $G_{00} =  3H^2  + 6H\partial _t \psi  - \frac{2}{{a^2 }}\nabla ^2 \psi$
\item $G_{0i}   =  2(H\partial _i \phi  - \partial _t \partial _i \psi )$
\item $G_{xx}   =  (\mathop a\limits^. \mathop {}\nolimits^2  + 2a\mathop a\limits^{..} )[ - 1 + 2(\phi  - \psi )] + (\partial _y \mathop {}\nolimits^2  + \partial _z \mathop{}\nolimits^2 )(\phi  + \psi ) - 2a^2 \partial _t \mathop {}\nolimits^2 \psi  + 2a\mathop a\limits^. \partial _t (\phi  - 3\psi )$
\item $G_{ij}   =   - \partial _i \partial _j (\phi  + \psi ) \ for \ i \neq j$ 
\end{itemize}

\paragraph{Aether field.}

We take a homogenous and spatially isotropic universe for the background, so the Aether must, in the background, respect this isotropy for the modified Einstein equations to have solutions, so only the time component can be non zero. The constraint on the norm is $g_{\alpha \beta} A^{\alpha} A^{\beta}=-1$ so in the background, we take $A^{\alpha}=\delta^{\alpha}_0$ and one can then expand it and write: 
\begin{equation}
A^\alpha   = \delta _0^\alpha   + \varepsilon B^\alpha  
\end{equation}

The constraint on the total vector fixes $B^0  =  - \phi$ with the perturbed form of the metric. \\

We can also derive:
\begin{equation}
\nabla A = \varepsilon \Sigma  + O(\varepsilon ^2 )
\end{equation}

with:
\begin{itemize}
\item $\Sigma _{i0}  =  - \frac{{\mathop a\limits^. }}{a}B_i $ 
\item $\Sigma _{0i}  = \partial _i \phi  + \partial _t B_i  - HB_i $ 
\item $\Sigma _{ij}  = \left[ {a\mathop a\limits^.  + a^2 \partial _t \psi  +  a\mathop a\limits^. (2\psi  - \phi )} \right]\delta _{ij}  + \partial _i B_j $
\end{itemize}

The other components vanish.

\paragraph{Matter.}

For matter fields, we can take: 
\begin{equation}
T_{\mu \nu } ^{matter}  = (\rho  + P)u_\mu  u_\nu   + Pg_{\mu \nu } 
\end{equation}

which is the stress-tensor of a perfect fluid without any anisotropic stress, with a density $\rho$, a pressure $P$ and with $u_  \mu$ the fluid four-velocity satisfying $g_{\mu \nu} u^ \mu u^ \nu = -1$. If we consider a non-relativistic fluid, ie with no spatial components of $u_ \mu$,  for this metric: $u_{ \mu } = (-1- \epsilon \phi, 0,0,0)$. \\

With this metric we have thus:
\begin{eqnarray}
 T_{00}^{matter}  = \left( {1 + 2\phi } \right)\rho  \\ 
 T_{0i}^{matter}  = T_{ij}^{matter}  = 0 \\ 
 T_{ii}^{matter}  = a^2 \left( {1 + 2\psi } \right)P 
 \end{eqnarray}

I carried out the calculations of the Einstein equations up to linear order analytically with this metric, except for the cross terms, and for the calculations are very tedious, time-consuming, and it is really easy to make mistakes, I tried to compute them with a calculus software, to check them and to calculate the cross-terms. I tried lots of Mathematica packages ("tensorial", "GREAT", "GRTesting", "xtensor") before finding the Maple "tensor" package that is a free one which can really carry out calculations with tensors, especially in which one can really enter components of a tensor (the Aether here) and that can compute covariant derivatives of it. I thus wrote the Maple sheet that is in the Annex B at the end of this report. 

\subsubsection{Static limit}

In the static limit,  the spatial terms of the Aether appear only at second order in all the equations.\\

$T_{\alpha \beta} ^{Aether}$ has no cross-terms (up to linear order), so we find:
\begin{equation}
\left.
\begin{array}{ll}
G_{ij}  =  8\pi GT_{ij}^{matter} + T_{ij}^{Aether} \\
G_{ij}  =  - \partial _i \partial _j (\phi  + \psi ) \ for \ i \neq j \\
T_{ij}^{Aether}=T_{ij}^{matter}=0  \ for \ i \neq j \\
\end{array}
\right\} \Rightarrow \phi = - \psi
\end{equation}

So the only non-zero component of the Einstein tensor is:
\begin{equation}
G_{00}  = 2 \nabla  {}^2 \phi 
\end{equation}

We have:
\begin{equation}
T_{00}^{Aether}  =  \left( {c_4  - c_1 } \right)\nabla  .\left( {F' \nabla   \phi } \right) - \frac{1}{2}\left( {1 + 2\varepsilon \phi } \right)FM^2 
\end{equation}

so we find the equation
\begin{equation}
\nabla  .\left( {\left( {2 + \left( {c_1  - c_4 } \right)F'} \right)\nabla   \phi } \right) = 8\pi G\rho 
\end{equation}

It looks tempting to recover MOND with such a Poisson equation including an extra term. \\

The kinetic scalar $K$ is in this case, up to second order:
\begin{equation}
K= \frac{c_4  - c_1}{{M^2 }} \left( {\nabla \phi } \right)^2
\end{equation}

Restricting to the case of $c_4=0$ considered by Zlosnik et al. (\cite{12}), we see that we can recover a MONDian regime in which $\nabla .\left( {\left| {\nabla \phi } \right|\nabla \phi } \right) \propto \rho$ in the limit of small $\left| {\nabla \phi } \right|$ writing $\mathop {\lim }\limits_{\left| {\nabla \phi } \right| \ll M} [2 + \left( {c_1 } \right)F'] \propto K^{\frac{1}{2}}$, so we can get a real MONDian limit if we consider $M$ of the same ordre of $a_0$.  \footnote{This was done by Zlosnik et al. (see \cite{12}). With no $c_4$, $K$ is ensured to be positive for consistence of a quantisized theory of the Aether carried out by Lim in \cite{10}. Including the $c_4$ safely would require further examination.}  More details and discussions are given in \cite{12}.

\subsubsection{Homogenous and isotropic universe}

The only non-zero component of the Einstein tensor are:
\begin{itemize}
\item $G_{00}=3H^2$
\item $G_{xx}  = G_{yy} = G_{zz} = - (\mathop a\limits^. \mathop {}\nolimits^2  + 2a\mathop a\limits^{..} )$
\end{itemize}

We find:
\begin{equation}
T_{00}^{Aether}  = 3c_2 \left( {2F'H^2  + \mathop {F'}\limits^. H + F'\frac{{\mathop a\limits^{..} }}{a}} \right)  
\end{equation}

The vector equations gives the Lagrange multiplier:
\begin{equation}
\lambda  = 3\left( {c_1  + c_2  + c_3 } \right)F'\frac{{\mathop a\limits^{..} }}{a} - 3c_2 \left( {\mathop {F'}\limits^. H + F'\frac{{\mathop a\limits^{..} }}{a}} \right)
\end{equation}

so that we get:
\begin{equation}
T_{00}^{Aether}  = 3\alpha F'\frac{{\mathop a\limits^{..} }}{a} - \frac{1}{2}M^2 F
\end{equation}

where $\alpha$ is defined as $\alpha=c_1+3 c_2+c_3$. \\

We have therefore the $00$ modified Einstein equation:
\begin{equation}
3\left( {1 - \alpha F'} \right)H^2  + \frac{1}{2}M^2 F = 8\pi G\rho 
\end{equation}

We can also calculate:
\begin{equation}
T_{xx}^{Aether}  = T_{yy}^{Aether}  = T_{zz}^{Aether}  =  - \alpha F' \left( 2\mathop a\limits^{. ^2}  + a\mathop a\limits^{..}  \right) - \alpha \mathop {F'}\limits^. a\mathop a\limits^. + \frac{1}{2}FM^2 
\end{equation}

and so we get the modified pressure equation:
\begin{equation}
 - \left( {1 - 2\alpha F'} \right)H^2  - 2\left( {1 - \frac{1}{2}\alpha F'} \right)\frac{{\mathop a\limits^{..} }}{a} + \alpha \mathop {F'}\limits^. H - \frac{1}{2}FM^2  = 8\pi GP
\end{equation}

Zlosnik et al. identified the additional terms that appear here with a cosmological constant. More details and discussions are once again present in \cite{12}.

\newpage

\subsubsection{General perturbed metric}

\paragraph{Aether with no spatial terms}

We consider first the case where the only non-vanishing component of $A$ is $A_0$, the spatial components being zero. 

\subparagraph*{Kinetic scalar}

We get for $K$:
\begin{eqnarray}
K = \frac{{3\alpha H^2 }}{{M^2 }} + \frac{{6\alpha H }}{{M^2 }}\left( { - H \phi  + \partial _t \psi } \right) + \frac{1}{{M^2 }}\frac{{(c_4  - c_1 )}}{{a^2 }}\left( {\nabla \phi } \right)^2  \nonumber \\
+ \frac{{3\alpha }}{{M^2 }}\left( { - 4H\psi \partial _t \psi  + 5H^2 \phi ^2  - 4H\phi \partial _t \psi  + \left( {\partial _t \psi } \right)^2 } \right)+\frac{{6c_2 H}}{{M^2 }}\partial _t \phi 
\end{eqnarray}

\subparagraph*{Density and pressure equations}

The $T_{00}$ component of the Aether stress-energy tensor is:
\begin{eqnarray}
T_{00} ^{Aether}  = 
 \frac{{(c_3  + c_4 - c_1 )}}{{a^2 }}\nabla .(F'\nabla \phi ) & \ & \; \; \; \; \;  \nonumber \\
+ 3c_2 \left[ {F'(2H^2  +  \frac{{\mathop a\limits^{..} }}{a} + 6H\partial _t \psi  - H\partial _t \phi  + \partial _t \mathop {}\nolimits^2 \psi ) + \mathop {F'}\limits^. (H + \partial _t \psi )} \right]  - \frac{1}{2}(1 + 2\phi )M^2 F + \lambda A_0 A_0 \nonumber
\end{eqnarray}
\begin{equation}
\end{equation}

The vector equation gives:
\begin{eqnarray}
 \lambda  = 3\left( {c_1  + c_2  + c_3 } \right)F'H^2  - 3c_2 \left( {F'\frac{{\mathop a\limits^{..} }}{a} + \mathop {F'}\limits^. H} \right) - \frac{{c_3 }}{{a^2 }}\partial _i \left( {F'\partial _i \phi } \right) + 3\left( {2\left( {c_1  + c_3 } \right)F'H - c_2 \mathop {F'}\limits^. } \right)\partial _t \psi  \nonumber \\
+ 3c_2 F'H\partial _t \phi  - 3c_2 F'\partial _t^2 \psi  + 6\left( {c_2 \left( {F'\frac{{\mathop a\limits^{..} }}{a} + \mathop {F'}\limits^. H} \right) + \left( {c_1  + c_2  + c_3 } \right)F'H^2 } \right)\phi \nonumber 
 \end{eqnarray}
\begin{equation}
\end{equation}

Hence we have:
\begin{equation}
T_{00}^{Aether}  =  - \frac{{c_1 - c_4 }}{{a^2 }}\nabla .(F'\nabla \phi ) + 3\alpha F'H^2  + 6\alpha F'H\partial _t \psi  - \frac{1}{2}(1 + 2\phi )M^2 F
\end{equation}

and thus:
\begin{equation}
 3(1 - \alpha F')H^2 - \frac{2}{{a^2 }}\nabla ^2 \psi  + \frac{{c_1 -c_4 }}{{a^2 }}\nabla .(F'\nabla \phi ) +  6 \left( 1 - \alpha  F' \right)  H\partial _t \psi  + \frac{1}{2}(1 + 2\phi )M^2 F = 8\pi GT_{00}^{matter}
\end{equation}

We have  $T_{00}^{matter}  = \left( {1 + 2\phi } \right)\rho$, so we get the modified density equation:
\begin{equation}
 3(1 - \alpha F')H^2  - \frac{2}{{a^2 }}\nabla ^2 \psi  + \frac{{c_1 - c_4 }}{{a^2 }}\nabla .(F'\nabla \phi ) + 6 \left( 1 - \alpha F' \right) H\partial _t \psi - 6 \left( 1 - \alpha F' \right) H^2 \phi + \frac{1}{2}M^2 F = 8\pi G\rho 
\end{equation}

\newpage

The spatial diagonal terms are equal and we have for instance:
\begin{eqnarray}
T_{xx}^{Aether}= \alpha ( - F' \left( 2\mathop a\limits^{. ^2}  + a\mathop a\limits^{..}  \right) - \mathop {F'}\limits^. a\mathop a\limits^. ) + \frac{1}{2} a^2 (1+2 \psi )  FM^2 \nonumber \\
+\alpha  \left[ {  - 3\left( {2F'a\mathop a\limits^.  + \mathop {F'a^2 }\limits^. } \right)\partial _t \psi  + F'a\mathop a\limits^. \partial _t \phi  - F'a^2 \partial _t^2 \psi  + 2\left( {F'\left( { {2\mathop a\limits^.} ^2  + a\mathop a\limits^{..} } \right) + \mathop {F'}\limits^. a \mathop a\limits^. } \right)\left( {\phi  - \psi   } \right)} \right] \nonumber
\end{eqnarray}
\begin{equation}
\end{equation}

We find therefore, as for matter,  $T_{ii}^{matter}  = a^2 \left( {1 + 2\psi } \right)P$, the modified pressure equation:
\begin{eqnarray}
 - \left( {1 - 2\alpha F'} \right)H^2  - 2\left( {1 - \frac{1}{2}\alpha F'} \right)\frac{{\mathop a\limits^{..} }}{a} + \alpha \mathop {F'}\limits^. H - \frac{1}{2}M^2 F + \frac{2}{{3a^2 }}{\nabla } ^2 \left( {\phi  + \psi } \right) \nonumber \\
+ \left( { - 6H\left( {1 - \alpha F'} \right) + \alpha \mathop {F'}\limits^. } \right)\partial _t \psi  + \left( {2 - \alpha F'} \right)H\partial _t \phi  - \left( {2 - \alpha F'} \right)\partial _t^2 \psi  \nonumber \\
+ \left( { - 2\alpha \mathop {F'}\limits^. H + 4\left( {1 - \alpha F'} \right)\frac{{\mathop a\limits^{..} }}{a} + 2\left( {1 - 2\alpha F'} \right)H^2 } \right)\phi  = 8\pi GP 
\end{eqnarray}

\subparagraph*{Cross terms}

We have
\begin{equation}
T_{0i}^{Aether}  = \left( {c_4  - c_1 } \right)\left( {F'\partial _t \partial _i \phi  + \left( {\mathop {F'}\limits^.  + HF'} \right)\partial _i \phi } \right)
\end{equation}

Thus we have, since the $T_{0i}^{matter}$ is taken to be zero here (for a non-relativistic perfect fluid):
\begin{equation}
\left( { - 2 + \left( {c_1  - c_4 } \right)F'} \right)\partial _t \partial _i \psi  + \left[ {\left( {2 + \left( {c_1  - c_4 } \right)F'} \right)H + \left( {c_1  - c_4 } \right)\mathop {F'}\limits^. } \right]\partial _i \phi  = 0
\end{equation}

We have for $i \neq j$, up to linear order 
\begin{equation}
T_{ij}^{Aether}  = 0
\end{equation}

So since the stress-tensor of matter has no cross terms either in the case considered, given the form of the Einstein-tensor cross terms $G_{ij}   =   - \partial _i \partial _j (\phi  + \psi ) \ for \ i \neq j$, we can, as in the static case, identify the potentials $\phi$ and $\psi$.\\

\paragraph{Aether with spatial terms}

We consider now also spatial terms: we take more precisely a covariant Aether of the form $(-1- \epsilon \phi, \epsilon B_x, \epsilon B_y, \epsilon B_z)$.  

\subparagraph*{Kinetic scalar}

We have, up to linear order:
\begin{equation}
K =\frac{{3\alpha H^2 }}{{M^2 }} + \frac{{6\alpha H }}{{M^2 }}\left( { - H \phi  + \partial _t \psi } \right) + 2\alpha \frac{{\mathop a\limits^. }}{{a^3 }}\partial _i B_i 
\end{equation}

The expression of $K$ to second order is very long and not very useful such. It is in Annex A.

\subparagraph*{Density and pressure equations}

We can derive:
\begin{equation}
T_{00}^{Aether}  =  - \frac{{c_1 - c_4 }}{{a^2 }}(\nabla .(F'\nabla \phi )+\partial _i F' \partial _t B_i  ) + 3\alpha F'H^2  + 6\alpha F'H\partial _t \psi  + 2\alpha \frac{{\mathop a\limits^. }}{{a^3 }}F' \partial _i B_i   - \frac{1}{2}(1 + 2\phi )M^2 F
\end{equation}

The expression of the Lagrange multiplier $\lambda$ used to get this is in Annex A.\\

We have thus:
\begin{eqnarray}
3(1 - \alpha F')H^2 - \frac{2}{{a^2 }}\nabla ^2 \psi  + \frac{{c_1 - c_4}}{{a^2 }}(\nabla .(F'\nabla \phi ) +\partial_i F' \partial _t B_i )   + 6 (1 - \alpha F') H\partial _t \psi   \nonumber \\
- 2\alpha \frac{{\mathop a\limits^. }}{{a^3 }}F'\partial _i B_i + \frac{1}{2}(1 + 2\phi )M^2 F = 8\pi GT_{00} 
\end{eqnarray}

and:
\begin{eqnarray}
3(1 - \alpha F')H^2 - \frac{2}{{a^2 }}\nabla ^2 \psi  + \frac{{c_1 - c_4 }}{{a^2 }} (\nabla .(F'\nabla \phi ) +\partial _i F'.\partial _t  B_i)  - 6 ( 1 - \alpha F') H^2 \phi  \nonumber \\
+ 6 ( 1 - \alpha F' ) H\partial _t \psi  - 2\alpha \frac{{\mathop a\limits^. }}{{a^3 }}F'\partial _i B_i + \frac{1}{2}M^2 F = 8\pi G\rho 
\end{eqnarray}

We find:
\begin{eqnarray}
T_{xx}^{Aether} =  \alpha ( - F' \left( 2\mathop a\limits^{. ^2}  + a\mathop a\limits^{..}  \right) - \mathop {F'}\limits^. a\mathop a\limits^. ) + \frac{1}{2} a^2 (1+2 \psi ) FM^2    \nonumber \\
+\alpha  \left[ {  - 3\left( {2F'a\mathop a\limits^.  + \mathop {F'a^2 }\limits^. } \right)\partial _t \psi  + F'a\mathop a\limits^. \partial _t \phi  - F'a^2 \partial _t^2 \psi  + 2\left( {F'\left( { {2\mathop a\limits^.} ^2  + a\mathop a\limits^{..} } \right) + \mathop {F'}\limits^. a \mathop a\limits^. } \right)\left( {\phi  - \psi   } \right)} \right] \nonumber \\
- \left( {c_1  + c_3 } \right)\left( {\mathop {F'}\limits^.  + F'\mathop a\limits^.} \right)\partial _x B_x  - \left( {c_2 \mathop {F'}\limits^.  + \left( {c_1  + 4c_2  + c_3 } \right)F'\mathop a\limits^.} \right)\partial _i B_i  - \alpha \mathop a\limits^. \partial _i F'B_i  \nonumber \\
- \left( {c_1  + c_3 } \right)  F'  \partial _t \partial _x B_x  - c_2 F'  \partial _t \partial _i B_i  \nonumber 
\end{eqnarray}
\begin{equation}
\end{equation}

The pressure equation becomes:
\begin{eqnarray}
 - \left( {1 - 2\alpha F'} \right)H^2  - 2\left( {1 - \frac{1}{2}\alpha F'} \right)\frac{{\mathop a\limits^{..} }}{a} + \alpha \mathop {F'}\limits^. H - \frac{1}{2}M^2 F + \frac{2}{{3a^2 }}{\mathop \nabla \limits^ \to } ^2 \left( {\phi  + \psi } \right) \nonumber \\
+ \left( { - 6H\left( {1 - \alpha F'} \right) + \alpha \mathop {F'}\limits^. } \right)\partial _t \psi  + \left( {2 - \alpha F'} \right)H\partial _t \phi  - \left( {2 - \alpha F'} \right)\partial _t^2 \psi  \nonumber \\
+ \left( { - 2\alpha \mathop {F'}\limits^. H +  + 4\left( {1 - \alpha F'} \right)\frac{{\mathop a\limits^{..} }}{a} + 2\left( {1 - 2\alpha F'} \right)H^2 } \right)\phi  \nonumber \\
 + \frac{\alpha }{{a^2 }}\left( {\frac{{\mathop {F'}\limits^.  + 4F'\mathop a\limits^. }}{3}\partial _i B_i  + \mathop a\limits^. \partial _i F'B_i  + \frac{1}{3}F'\partial _t \partial _i B_i } \right) = 8\pi GP 
\end{eqnarray}

\subparagraph*{Cross terms}

We find:
\begin{eqnarray}
T_{0x}  = \left( {c_4  - c_1 } \right)\left( {F'\partial _t \partial _x \phi  + \left( {\mathop {F'}\limits^.  + HF'} \right)\partial _x \phi } \right) \nonumber \\
\left( {c_4  - c_1 } \right)\left( {F'\partial _t \mathop {}\nolimits^2 B_x  + (\mathop {F'}\limits^. + F'H) \partial _t B_x \partial _t B_x } \right) + \alpha \left( {F'\frac{{\mathop a\limits^{..} }}{a} + \mathop  {F'}\limits^. H  - F'H^2 } \right) B_x  \nonumber \\
+\frac{{c_3  - c_1 }}{{2a^2 }}\left( {\partial _i F'\left( {\partial _x B_i  - \partial _i B_x } \right) + F'\partial _i \left( {\partial _x B_i  - \partial _i B_x } \right)} \right)
\end{eqnarray}

and so we have:
\begin{eqnarray}
\left( { - 2 + \left( {c_1  - c_4 } \right)F'} \right)\partial _t \partial _x \psi  + \left[ {\left( {2 + \left( {c_1  - c_4 } \right)F'} \right)H + \left( {c_1  - c_4 } \right)\mathop {F'}\limits^. } \right]\partial _x \phi  \nonumber \\
-\left( {c_4  - c_1 } \right)\left( {F'\partial _t \partial _x \phi  + \left( {\mathop {F'}\limits^.  + HF'} \right)\partial _x \phi } \right) \nonumber \\
\left( {c_4  - c_1 } \right)\left( {F'\partial _t \mathop {}\nolimits^2 B_x  + (\mathop {F'}\limits^. + F'H) \partial _t B_x \partial _t B_x } \right) + \alpha \left( {F'\frac{{\mathop a\limits^{..} }}{a} + \mathop  {F'}\limits^. H  - F'H^2  } \right) B_x  \nonumber \\
+\frac{{c_3  - c_1 }}{{2a^2 }}\left( {\partial _i F'\left( {\partial _x B_i  - \partial _i B_x } \right) + F'\partial _i \left( {\partial _x B_i  - \partial _i B_x } \right)} \right)=0
\end{eqnarray}

We find that the Aether stress-energy tensor has now spatial cross terms in linear order. For $i \neq j$, we have: \\
\begin{equation}
T_{ij}^{Aether}  =  - \frac{c_1  + c_3}{2} (F' H \partial _{(i} B_{j)}  + F'\partial _t \partial _{(i} B_{j)}  + \mathop {F'}\limits^. \partial _{(i} B_{j)} )
\end{equation}

We see thus that the spatial terms of the Aether makes the two scalar potentials differ from one another. We cannot therefore identify them as we could do for an Aether without spatial terms. Notice that in GR, these two potentials can only be made to differ by off-diagonal terms of the matter stress-energy tensor, ie by anisotropic stress. One can also see that $T_{ij}$ is zero if $c_1=-c_3$, in the magnetic case.\\

I have thus got the perturbations of the Aether stress-energy tensor and the Einstein equation for an Aether with or without spatial terms and a Lagrangian involving a general function of the kinetic term $K$. \\

These equations are very general. Simplifying assumptions must be made for physics to arise. I unfortunately did not have time to go on much with this during my internship but work is currently being done about it, keeping this general perturbed metric, but considering for instance the vector perturbations as gradient of a scalar field. For instance, in this particular case, and considering $F'$ as constant, the the $0x$ equation, would reduce to the one of a harmonic oscillator, with a negative string contant and a source term which is a function of the scalar potentials $\phi$ and $\psi$. One can therefore explore (numerically) the growing of this scalar potential from an initial instability. Similar work was done by Dodelson and Liguori (see \cite{19}) , but for the Bekenstein TeVeS Lagrangian. The generality of the equations presented here gives nevertheless the opportunity of exploring various cases just by simplifying them.

\newpage

\subsection{Scalar field}

We can try to take $F(K)=K$ and make the coefficients of the Lagrangian depend on a scalar field "instead". Similar work has been done by Kanno and Soda (\cite{18}, \cite{20}) in the context of cosmic inflation.\\

With:
\begin{equation}
S = \int {d^4 x\sqrt { - g} \left[ {\frac{R}{{16\pi G_N }} + L(A,g,\chi)} \right]}  + S_M 
\end{equation}

and:
\begin{equation}
L(A,g,\chi) = \frac{{1 }}{{16\pi G_N}} \left( M^2 K + \lambda (A^\alpha  A_\alpha   + 1)  - V\left( \chi  \right) - \frac{{\left( {\nabla \chi } \right)^2 }}{2} \right)
\end{equation}

the scalar equation is:
\begin{equation}
 - \frac{{dV}}{{d\chi }} + \frac{{dc_1 }}{{d\chi }}\nabla _\alpha  A^\gamma  \nabla ^\alpha  A_\gamma   + \frac{{dc_2 }}{{d\chi }}\left( {\nabla _\alpha  A^\alpha  } \right)^2  + \frac{{dc_3 }}{{d\chi }}\nabla _\alpha  A^\gamma  \nabla _\gamma  A^\alpha   + \frac{{dc_4 }}{{d\chi }}A^\alpha  A^\beta  \nabla _\alpha  A_\gamma  \nabla _\beta  A^\gamma  =  - \nabla _\nu  \nabla ^\nu  \chi 
\end{equation}

and the Einstein equation, considering $g^{\alpha \beta}$, $A^{mu}$ and the scalar field $\phi$ as three degrees of freedom is:
\begin{equation}
G_{\alpha \beta }  =  T_{\alpha \beta }^{Aether} + T_{\alpha \beta }^{scalar} + 8\pi GT_{\alpha \beta }^{matter} 
\end{equation}

I defined here (in the action) the scalar tensor by analogy with the Aether tensor as far as the coefficients in the Einstein equations are concerned. 

We have:
\begin{equation}
T_{\alpha \beta }^{scalar}  =  - \frac{g_{\alpha \beta }}{2} \left( {\frac{{\left( {\nabla \chi } \right)^2 }}{2} + V(\chi )} \right) + \frac{1}{2} \nabla _\alpha  \chi \nabla _\beta  \chi 
\end{equation}

\paragraph{Cosmology}

We consider the case of a homogeneous and isotropic universe with: $ds^2  =  - dt^2  + a(t)^2 (dx^2  + dy^2  + dz^2 )$. 
The scalar field can depend only on time. (If we make it depend on spatial coordinates, the scalar stress-energy tensor will have non-cross terms $T_{ij}^{scalar}  =\frac{1}{2} \left( {\partial _i \chi } \right)\left( {\partial _j \chi } \right)$ for $i \neq j$ that are zero using the Einstein equations.)

The only non-vanishing terms of the scalar stress-energy tensor are thus the diagonal ones, which are:
\begin{eqnarray}
 T_{00}^{scalar}  = \frac{1}{4}\left( {\partial _t \chi } \right)^2  +  \frac{1}{2} V\left( \chi  \right) \\ 
 T_{xx}^{scalar}  = T_{yy}^{scalar}  = T_{zz}^{scalar}  =  - \frac{1}{4}a^2 \left( { - \left( {\partial _t \chi } \right)^2  + 2V\left( \chi  \right)} \right) 
 \end{eqnarray}

The scalar equation is:
\begin{equation}
 - \frac{{dV}}{{d\chi }} + 3H^2 \frac{{d\alpha }}{{d\chi }} = \partial _t^2 \chi  + 3H\partial _t \chi
\end{equation}

with $\alpha \left( \chi \right)  = c_1 \left( \chi \right) + 3c_2 \left( \chi \right) + c_3 \left( \chi \right) $.

and the Einstein equations are:
\begin{eqnarray}
3\left( {1 - \frac{\alpha \left( \chi  \right)}{2}} \right)H^2  = \frac{1}{4}\left( {\partial _t \chi } \right)^2  +  \frac{1}{2} V\left( \chi  \right) + 8\pi G\rho \\
 - \left( {1 - \frac{\alpha \left( \chi \right) }{2} } \right)H^2  - 2\left( {1 - \frac{1}{2}\alpha \left( \chi \right) } \right)\frac{{\mathop a\limits^{..} }}{a}  =  \frac{1}{4} \left( {\partial _t \chi } \right)^2  -\frac{1}{2} V\left( \chi  \right) + 8\pi GP
\end{eqnarray}

If we try to take a constant scalar field and a simple scalar potential $V\left( \chi  \right) = \frac{1}{2}M^2 \chi ^2 $, the scalar equation is so:
\begin{equation}
 - M^2 \chi  + 3H^2 \frac{{d\alpha }}{{d\chi }} =0
\end{equation}

and the Einstein equations are:
\begin{eqnarray}
3\left( {1 - \frac{\alpha \left( \chi  \right)}{2}} \right)H^2  = \frac{1}{4}M^2 \chi ^2   + 8\pi G\rho \\
 - \left( {1 - \frac{\alpha \left( \chi \right) }{2} } \right)H^2  - 2\left( {1 - \frac{1}{2}\alpha \left( \chi \right) } \right)\frac{{\mathop a\limits^{..} }}{a}  = -\frac{1}{4}M^2 \chi ^2 + 8\pi GP
\end{eqnarray}

or, using the scalar equation:
\begin{eqnarray}
3\left( {1 - \frac{\alpha \left( \chi  \right)}{2}} \right)H^2  - \frac{{9H^4 }}{{2M^2 }}\left( {\frac{{d\alpha }}{{d\chi }}} \right)^2  = 8\pi G\rho \\
 - \left( {1 - \frac{\alpha \left( \chi \right) }{2} } \right)H^2  - 2\left( {1 - \frac{1}{2}\alpha \left( \chi \right) } \right)\frac{{\mathop a\limits^{..} }}{a}  +  \frac{{9H^4 }}{{2M^2 }}\left( {\frac{{d\alpha }}{{d\chi }}} \right)^2= 8\pi GP
\end{eqnarray}

We can try to identify the additionnal terms in the modified Einstein equations with a cosmological constant $\Lambda$ (like what can be done for a simple Aether, with no scalar field \cite{12}) ie such that $G_{\alpha \beta} + \Lambda g_{\alpha \beta} = 8 \pi G  T_{\alpha \beta} $, like what was done in Zlosnik. We see that we should have:
\begin{equation}
\left( {\frac{{d\alpha }}{{d\chi }}} \right)^2 +  \frac{{M^2}}{{3H^2}} \  \alpha \left( \chi  \right) = \frac{{2M^2}}{{9H^4}}  \Lambda 
\end{equation}

We can't therefore do it for $H$ that is time-dependent appears in the coefficients of the differential equation whereas $\alpha$ is a function of the sole scalar field which is taken constant here.

\newpage

\subsection{Maple code}

The Maple sheet I wrote aims at computing the equations of the report with a large choice of parameters. It follows the exact notations (indices, etc) and the exact steps of the report. Maple can't carry out the calculations from the Lagrangian, it can't variate the action w.r.t. the metric or the Aether, so the tensorial equations after differenciation must be entered in it.\\  

The sheet of the Annex is expanded for a simple FRW metric because it is the metric that gives the lightest Maple results. These results can, in spite of collection of terms by Maple, be several pages long for a FRW perturbed metric and an Aether with spatial terms. 

\paragraph{The first section} allows to choose a system of coordinates and a metric. Five types of metric are displayed in the annex: a simple FRW metric, a perturbed metric with static potentials, the FRW perturbed metric of the report, the same in conformal time (conformal time $\tau$ is sometimes used in papers, it is defined by $a(\tau)^2 d \tau^2 = dt^2$, and a spherically symetric metric which I added to the Maple code but did not mention in my report. One can add off-diagonal terms easily in a metric, for all components are defined separatly. All tools to compute the Einstein tensor, and the Einstein tensor are computed (Those are procedures included in the package). The Aether field is then defined, including spatial terms or not (when it is consistent with the chosen metric, ie not for a FRW metric for which spatial terms will break the isotropy). 

\paragraph{The second session} aims at computing the Aether stress-energy tensor. All intermediate tensors are defined so that one can change them individually. The tensor $K^{\alpha \beta}_{\gamma \sigma}$ can be easily changed if one wants for instance to put no constraint on the norm of the Aether and thus to add terms in it. One can choose to take a general function $F$ of the kinetic term $K$, to take $F(K)=K$, to have a "simple" Aether theory, to take a Maxwell-like $K$ to simplify even more, and one could enter a specific function. Some components are expanded at the end of the section. Notice that the input lines are very long because the results given by Maple are often messy, so one must arrange them by collecting terms (in the Annex, some commands used for other metrics are left in the input lines).

\paragraph{The third section} computes the stress-energy tensor (the one of a perfect fluid) for matter, given the four-velocity defined in the first section.

\paragraph{The fourth section} is optionnal. I wrote it to include a scalar field whom the coefficients of the Aether Lagrangian depend on (see last section). It computes the scalar stress-energy tensor and the scalar equation.

\paragraph{The fifth section} computes the modified Einstein equations. \\

This calculus sheet was approved by Tom Zlosnik that works on the Aether in the University of Oxford. (I sent it to him but did not collaborate further with him.)

\section{The EFE: example of a galaxy disc with a bulge}

\textit{EFE: External Field Effect}\\

The MOND gravitationnal equation is such that if any external field is considered, it will break the Strong Equivalence Principle which reads that gravity can always be replaced by an accelerating frame. This is fundamental for GR, but once again, it is just a principle and might not be true. Nature might not be as simple as we often wish it is. The external field in MOND has effects on the interior dynamics of a system embedded in it. This should not be confused with tidal effects because this occurs even if the external field is uniform. \\

I studied an example of how a static external field can affect a system, a galaxy disk with a bulge around it. This real MOND problem involves various galactic dynamics subjects: potential issues, phase space densities, etc... My supervisor told me about his work on this problem: I TeXed it for him, as well as another two pages about phase space density considerations, explaining it with clarified notations, drawings, etc...

\subsection{Kuzmin disks and the Plummer model in Newtonian gravity}

The Plummer model was first used by Plummer to describe globular clusters, massive stellar systems containing $10^4-10^6$ stars in a nearly spherical distribution (see \cite{17}).
 
We consider the spherical potential:
\begin{equation}
\Phi _{Plummer}  =  - \frac{{GM}}{{\sqrt {r^2  + r_c^2 } }}
\end{equation}

$r_c$ is the core radius (the radius where the surface brightness has fallen to half its central value).

We find thus:
\begin{eqnarray}
\rho_{Plummer} \left( r \right) & = & \frac{1}{{4\pi Gr^2 }}\frac{d}{{dr}}\left( {r^2 \frac{{d\Phi _P }}{{dr}}} \right)  \\ 
& = & \frac{{3Mr_c^2 }}{{4\pi \left( {r^2  + r_c^2 } \right)^{\frac{5}{2}} }}  \\
& = & - \frac{{3Mr_c^2 }}{{4\pi \left( GM \right) ^5 }}\Phi _{Plummer} ^5  
\end{eqnarray}

To describe galaxy infinitely thin disks, Kuzmin introduced a potential pair, consisting of two fictive point masses.
We consider a disk and an axisymmetric potential given by:
\begin{equation}
\Phi _K  =  - \frac{{GM}}{{\sqrt {R^2  + \left( {\left| z \right| + z_0 } \right)^2 } }}
\end{equation}

\begin{figure}[h!]
\center
\includegraphics[height=5cm]{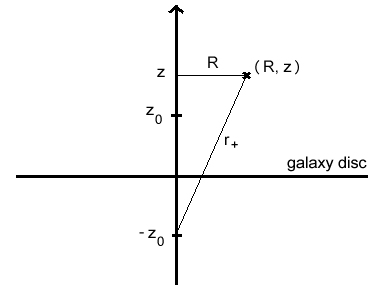} 
\caption{Kuzmin potential}
\end{figure}

We have therefore:
\begin{equation}
\Phi_K = \left\{
    \begin{array}{ll}
         - \frac{{GM}}{{r_ +  }} & \mbox{above the plane}  \\
         - \frac{{GM}}{{r_ -  }} & \mbox{below the plane} \\
         - \frac{{GM}}{{\sqrt {R^2  + z_0 ^2 } }} & \mbox{on the plane}
    \end{array}
\right.
\end{equation}

So above the plane, the potential is the one of a (fictive) point mass located at $-z_0$, and below the plane, the potential is the one created by a fictive point mass at $z_0$.\\

The force has therefore a $z$ component discontinuous on the plane of the disk (the $R$ one is continuous). The density is zero above and below the plane, because the potential is there "point-mass like", but one can apply the Gauss theorem to an infinitely thin box for instance with upper and lower parts right above and below the plane to calculate the surface density $\sigma$ of the disk:
\begin{eqnarray}
& & \frac{{\partial \Phi _K }}{{\partial z}}\left| {_{z = 0^ +  }  - } \right.\frac{{\partial \Phi _K }}{{\partial z}}\left| {_{z = 0^ -  }  = } \right.4\pi G\Sigma  \\
& & \Rightarrow \Sigma  = \frac{1}{{2\pi }}\frac{{Mz_0 }}{{\left( {R^2  + z_0 ^2 } \right)^{\frac{3}{2}} }} 
\end{eqnarray}

We have thus a mass-distribution which is purely in the disk.

If we want now to consider not only a thin disk, but a bulge as well, we can take now the potential:
\begin{equation}
\Phi   =  - \frac{{GM}}{{\sqrt {R^2 + r_c^2 + \left( {\left| z \right| + z_0 } \right)^2 } }}
\end{equation}

so that we get spherical Plummer potentials above and below the plane:
\begin{equation}
\Phi = \left\{
    \begin{array}{ll}
         - \frac{{GM}}{{\sqrt{r_c^2 + r_+ ^2}  }} & \mbox{above the plane}  \\
         - \frac{{GM}}{{\sqrt{r_c^2 + r_- ^2}  }} & \mbox{below the plane} \\
    \end{array}
\right.
\end{equation}

so we have a Plummer density above and below the plane.\\

We can once again use the discontinuity of the potential at $z=0$ to calculate the surface density of the disk, which depends now on $R$:
\begin{eqnarray}
\Sigma \left(R \right) = \frac{1}{{2\pi }}\frac{{Mz_0 }}{{\left( {r_c ^2 + R^2  + z_0 ^2 } \right)^{\frac{3}{2}} }} 
\end{eqnarray}

Its mass is:
\begin{eqnarray}
M_{disk}  & = & \int_0^\infty  {2\pi } \Sigma \left( R \right)RdR \\
 & = & \frac{{Mz_0 }}{{\sqrt {r_c ^2  + z_0 ^2 } }} 
\end{eqnarray}

and with $\rho_+$ the density above the disk, because of the symmetry, the mass of the bulge is:
\begin{eqnarray}
M_{bulge}  & = & 2\int_{R = 0}^\infty  {\int_{z = 0}^\infty  {2\pi \rho _ +  \left( {R,z} \right)RdRdz} }  \\
 &  = & M\left( {1 - \frac{{z_0 }}{{\sqrt {r_c ^2  + z_0 ^2 } }}} \right) 
\end{eqnarray}

\subsection{Kuzmin disk with a bulge in MOND}

\subsubsection{With no external field}

To carry out calculations, we assume the Bekenstein $\mu$ function ( \cite{7}) : 
\begin{eqnarray}
\mu \left( x \right) = \frac{{ - 1 + \sqrt {1 + 4x} }}{{1 + \sqrt {1 + 4x} } } \ \ with \ \ x = \frac{ \left| g \right| }{a_0} 
\end{eqnarray}

The MOND equation for a mere sphere is:
\begin{eqnarray}
 \nabla  .\left( {\mu \left( x \right)\nabla  \Phi } \right) = 4\pi G\rho  = \nabla  ^2 \Phi _N  \\ 
  \Rightarrow \nabla  .\left( {\mu \left( x \right) \nabla  \Phi  - \nabla  \Phi _N } \right) = 0 
\end{eqnarray}

In general, this thus implies:
\begin{equation}
\mu (x) \mathbf{g} = {\mathbf{g}} _N +  \nabla \times \mathbf{h}
\end{equation}

But for spherical symmetry, we have (one can see it by integrating and using the Stokes theorem):
\begin{equation}
\mu (x) \mathbf{g} = {\mathbf{g}} _N
\end{equation}

With the particular form of the $\mu$ function, we get:
\begin{equation}
\left| g \right| = \left| g_N \right| +\sqrt{ \left| g_N \right| a_0}
\end{equation}

We can consider a gravity center at the location $(0,- z_0)$, below the plane, and we have thus:
\begin{equation}
\mathbf{g} \left( {r_ +  } \right) =  - \left| {g_N \left( {r_ +  } \right)} \right|\mathbf{e} {{} _r} _+   - \sqrt {g_N \left( {r_ +  } \right)a_0 } \mathbf{e}{{} _r}_ +  
\end{equation}

The gravitationnal field $g_N \left( r_+ \right)$ is a Newtonian one, so a Kuzmin one here, ie:
\begin{eqnarray}
\mathbf{g} \mathop {}\nolimits_N \left( {r_ +  } \right)  & = & - \frac{d}{{d\mathbf{r_ +  }  }} \left(- \frac{{GM}}{{\sqrt {R^2 + r_c^2 + \left( {\left| z \right| + z_0 } \right)^2 } }} \right) \mathbf{e} {{} _r} _+ \\
& = &  - \frac{{GM \left[ {\mathbf{R}   + \left( {\mathbf{z}  + \mathbf{{z_0 }}  } \right)} \right] }}{{ \left( R^2 + r_c^2 + \left( {\left| z \right| + z_0 } \right)^2 \right)^{3/2}  }}  
\end{eqnarray}

and the gravitationnal field below the plane is the mirror image of this.\\

The potential itself is:
\begin{eqnarray}
\Phi \left( {r_ +  } \right) & = &  - \int_ {+\infty}^{r_+}  {{g\left( {r_ +  } \right)} dr_ +  }  \\
 & = & - \int_ {+ \infty} ^{r_+}  {\left( { {g_N \left( {r_ +  } \right)} - \sqrt {\left| {g_N \left( {r_ +  } \right)} \right|a_0 } } \right)dr_ +  }  \\
 & = & \Phi _{Newton}  + \Phi _{effective \ DM} 
\end{eqnarray}

\subsubsection{Sphere dominated by an external field}

We consider now a sphere dominated by an external field in a MOND gravity, ie :
\begin{eqnarray}
 \nabla   .\left( {\mu \left( {\frac{{\left| {\nabla   \Phi } \right|}}{{a_0 }}} \right)\nabla  \Phi } \right) & = & 4\pi G\rho  \\
 with \ - \nabla   \Phi  = \mathbf{g_{ext} }  - \nabla   \Phi _ {int}  & , & \ \left| {\mathbf{g_{ext} } } \right|  \gg  \left| {\nabla  \Phi _ {int} } \right| 
\end{eqnarray}

We can see that the "external field effect" of MOND is due to the non-linearity of this equation: we can expand $\mu$:
\begin{eqnarray} 
\mu & = &  \mu \left( {\frac{{\left| {\mathbf{g_{ext} }   - \nabla   \Phi _{{\mathop{\rm int}} } } \right|}}{{a_0 }}} \right)  \\ 
& = & \mu \left( {\frac{{\left| {\mathbf{g_{ext} }  } \right|}}{{a_0 }}} \right) - \left. {\frac{{d\mu }}{{dg}}} \right|_{\left| {\mathbf{g_{ext} } } \right|} \mathbf{g_{ext} }  .\nabla  
\Phi _{{\mathop{\rm int}} }   \\ 
\Rightarrow \mu & = & \mu _e \left( {1 - \frac{{L_e }}{{\left| {\mathbf{g_{ext} }  } \right|}}\frac{{\partial \Phi _{ int}}}{{ \partial Z}}  }\right) \ with \ 
\mu _e  = \mu \left( {\frac{{\left| {\mathbf{g_{ext} }  } \right|}}{{a_0 }}} \right) \ and \ L_e  = \left. {\frac{{d\ln \mu }}{{d\ln \left| {\mathbf{g}  } \right|}}} \right|_{\left| {\mathbf{g_{ext} }  } \right|} 
 \end{eqnarray}

and the MOND equation is thus:
\begin{equation}
\nabla   .\left[ {\mu _e \left( {1 - \frac{{L_e }}{{\left| {\mathbf{g_{ext} }  } \right|}}\frac{{\partial \Phi _{{\mathop{\rm int}} } }}{{\partial Z}}} \right)\left( { - \mathbf{g_{ext} }   + \nabla   \Phi _{{\mathop{\rm int}} } } \right)} \right] = 4\pi G\rho 
\end{equation}

so up to linear order in  $\frac{{\left| {\mathop \nabla \limits^ \to  \Phi _{{\mathop{\rm int}} } } \right|}}{{\left| {\mathop {g_{ext} }\limits^ \to  } \right|}}$, we have.
\begin{equation}
\mu _e \left[ { - \nabla   .\mathbf{g_{ext} }   + \nabla ^2 \Phi _{{\mathop{\rm int}} }  + L_e \frac{{\partial ^2 \Phi _{{\mathop{\rm int}} } }}{{\partial Z^2 }}} \right] = 4\pi G\rho 
\end{equation}

The first term is zero for the external field is taken uniform, so:
\begin{eqnarray} 
\mu _e \left[ {\nabla ^2  + L_e \frac{{\partial ^2 }}{{\partial Z^2 }}} \right]\Phi _{{\mathop{\rm int}} }  & = & 4\pi G\rho  \\
\left[ {\frac{{\partial ^2 }}{{\partial X^2 }} + \frac{{\partial ^2 }}{{\partial Y^2 }} + L_1 \frac{{\partial ^2 }}{{\partial Z^2 }}} \right]\Phi _{{\mathop{\rm int}} }  & = & 4\pi G\frac{\rho }{{\mu _e }}, \ with \ L_1=L_e+1 
\end{eqnarray}

For a Newtonian gravity (with $\mu=1$ and $L_e=0$) we would find the mere Poisson equation for the internal potential, whereas here, we find an equation similar to the Poisson equation, but anisotropic and for a higher density $\frac{\rho }{{\mu _e }}$.\\

With the transformation
\begin{equation}
X' = X ,\ Y' = Y, \ Z' = \frac{Z}{{\sqrt {L_1 } }}
\end{equation}

we find the Poisson equation in these new coordinates, still with the same higher density:
\begin{equation} 
\nabla ^2 _, \Phi _{{\mathop{\rm int}} }  =  4\pi G\frac{{\rho}}{{\mu _e }} 
\end{equation}

where the $,$ refers to the  $\left( {X',Y',Z'} \right)$ coordinates.

\subsubsection{Ellipsoid dominated by an external field}

We can consider the physical density: 
\begin{eqnarray}
\rho & = & \frac{1}{{\sqrt{L_1} }} \rho_{Plummer} \left( r' \right)  \\
 & = & \frac{1}{{\sqrt{L_1} }}\frac{{3Mr_c ^2}}{{4\pi \left( {r'^2  + r_c ^2 } \right)^{\frac{5}{2}} }} , \ with \ r' = \sqrt {X'^2  + Y'^2  + Z'^2} 
\end{eqnarray}

where the $\frac{1}{{\sqrt{L_1} }}$ factor is such as to get a total mass integrated over all space equal to $M$. This is the density of an ellipsoid (the surfaces of equidensity are surfaces such that $X^2  + Y^2  + \frac{Z^2}{{L_1}} =cste$, ie ellipsoids). We have the perturbed MOND equation:
\begin{eqnarray} 
\nabla ^2 _, \Phi _{{\mathop{\rm int}} }  & = & 4\pi G\frac{{\rho}}{{\mu _e }}  \\
 & = & 4\pi G\frac{{\rho_{Plummer} \left( {r'} \right)}}{{\sqrt{L_1} \mu _e }}  
\end{eqnarray}

So we have:

\begin{equation}
\Phi _{int}  \left( {r'} \right) = \frac{{\Phi _{Plummer} \left( {r'} \right)}}{{\sqrt{L_1}\mu _e }}
\end{equation}

The relation between the density $\rho$ and the potential $\phi_{int}$ differs therefore by a factor $ \mu _e^5 L_1^{5/2}$:
\begin{equation}
\rho  =  - \mu _e ^5 L_1 ^{\frac{5}{2}} \frac{{3Mr_c ^2 }}{{4\pi \left( {GM} \right)^5 }}\Phi _{int} ^5 
\end{equation}

If we assume that the origins of the elliptic potentials are in the $\left( {X',Y'} \right)$ plane, at $\left( {-X'_0,-Y'_0} \right)$  and $\left( {X'_0,Y'_0} \right)$, that the external field is also in this plane, and we define the coordinates $\left( {x,y,z} \right)$ such that the $\left( x,z \right)$ is the same as the $\left( X,Z \right)$ one and by requiring that the potential is continuous on a plane $z=0$, we have on this plane:
\begin{eqnarray} 
\Phi_{int} & = &  - \frac{{GM}}{{\sqrt{L_1} \mu _e }}\frac{1}{{\sqrt {r_c ^2  + {{r_ +}'} ^2 } }}  \\
& = & - \frac{{GM}}{{\sqrt{L_1} \mu _e }}\frac{1}{{\sqrt {r_c ^2  + {{r_ -}'}  ^2 } }}  \\
 \Rightarrow r'_ +   & = & r'_ -  \nonumber
\end{eqnarray}

so that we can write for a point on this plane: 
\begin{eqnarray} 
\left( {X' - X'_0 } \right)^2  + Y'^2  + \left( {Z' - Z'_0 } \right)^2  & = & \left( {X' + X'_0 } \right)^2  + Y'^2  + \left( {Z' + Z'_0 } \right)^2   \\
\Rightarrow \left( {X - X'_0 } \right)^2  + Y^2  + \left( {\frac{Z}{{\sqrt {L_1 } }} - Z'_0 } \right)^2  & = & \left( {X + X'_0 } \right)^2  + Y^2  + \left( {\frac{Z}{{\sqrt {L_1 } }} + Z'_0 } \right)^2   \\
\Rightarrow XX'_0  & = & - \frac{{ZZ'_0 }}{{\sqrt {L_1 } }} \nonumber
\end{eqnarray}

\begin{figure}[h!]
\center
\includegraphics[height=5cm]{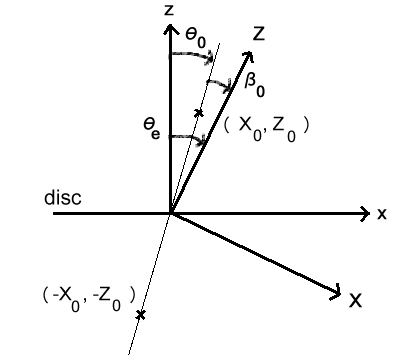} 
\caption{Angles and coordinates}
\end{figure}

With $\theta_e$ the angle between the z axis and the Z one (ie the one along wich the external field is), we have therefore:
\begin{eqnarray}
\tan \theta _e  & = & \sqrt {L_1 } \frac{{X'_0 }}{{Z'_0 }}  \\
 & = & L_1 \frac{{X_0 }}{{Z_0 }} 
\end{eqnarray}

If we define $\beta_0$ and $\theta_0$ such that:
\begin{eqnarray}
\tan \beta _0  & = & \frac{{X_0 }}{{Z_0 }}  \\
and \ \theta _0  & = & \theta _e  - \beta _0 
\end{eqnarray}

we find:
\begin{eqnarray}
\tan \theta _0  & = & \frac{{\tan \theta _e  - \tan \beta _0 }}{{1 + \tan \theta _e \tan \beta _0 }}  \\
 & = & \frac{{\left( {L_1  - 1} \right)\sin \theta _e \cos \theta _e }}{{L_1 \cos ^2 \theta _e  + \sin ^2 \theta _e }} 
\end{eqnarray}

So as we turn on an external field, we see that the center of the potentials shift and their axis rotates.\\

We have above the plane:
\begin{eqnarray}
\Phi _{int}  & = & - \frac{1}{{\mu _e  }}\frac{{GM}}{{\sqrt {L_1 r_c ^2  + A \left( {x + x_0 } \right)^2  + L_1 y^2  + B \left( {z + z_0 } \right)^2  + C \left( {x + x_0 } \right)\left( {z + z_0 } \right) } }} \\
with & & A  =  \left( {L_1 \cos ^2 \theta {}_e + \sin ^2 \theta _e } \right) , \\
& & B  =  \left( {L_1 \sin ^2 \theta _e  + \cos ^2 \theta {}_e} \right) , \\
& & C  =  2 \left( {L_1  - 1} \right) \cos \theta _e \sin \theta _e
\end{eqnarray}

and we can also find:
\begin{equation}
x_0  =  - z_0 \frac{{L\cos \theta _e \sin \theta _e }}{{1 + L\cos ^2 \theta _e }}
\end{equation}

\subsubsection{Potential energy}

I calculated the potential energy for the potential:
\begin{equation}
PE = \frac{1}{2}\iiint{\rho \phi _{{\mathop{\rm int}} } d^3 x}
\end{equation}

We find:
\begin{eqnarray}
 PE =  - \frac{{3GM^2 }}{{16\mu _e L_1 ^2 }}\left( {\frac{\pi }{{4r_c }} - \frac{{z_0 \sqrt A }}{{2r_c ^2 \left( {Ar_c ^2  + z_0 ^2 } \right)}} - \frac{1}{{2r_c }}\arctan \frac{{z_0 }}{{r_c \sqrt A }}} \right) \\ 
A = L_1 \cos ^2 \theta _e  + \sin ^2 \theta _e 
\end{eqnarray}

This sole result is not enlightning. With no external field, we would have just the first term. One could explore Virial theorem related questions, carry out some numerical simulations, etc. I did not go much further about this because I came back to the Aether.

\newpage

\section*{Conclusion}

During this internship, I discovered MOND, its birth, building, and got in touch with some MOND issues. I also discovered the Aether, which is a field by itself and is linked to a huge amount of various attempts of modifications of General Relativity. I rederived the fundamental equations of the theory, calculated them for different metrics and found a right way to compute them. I calculated them for a FRW perturbed metric, and a general function of the kinetic terms appearing in the Lagrangian, which had not been done such yet. I wrote a Maple code (approved by Tom Zlosnik of the University of Oxford) that can be used for various computations around the modified Einstein equations. This sheet is a useful tool to carry out calculations and can be easily modified. I also got interested in a problem of galactic dynamics in MOND, involving the special "external field effect". Some further work is currently being done about the Aether and the growth of perturbations. 

\vfill

\begin{figure}[h!]
\centering
\includegraphics[height=8cm]{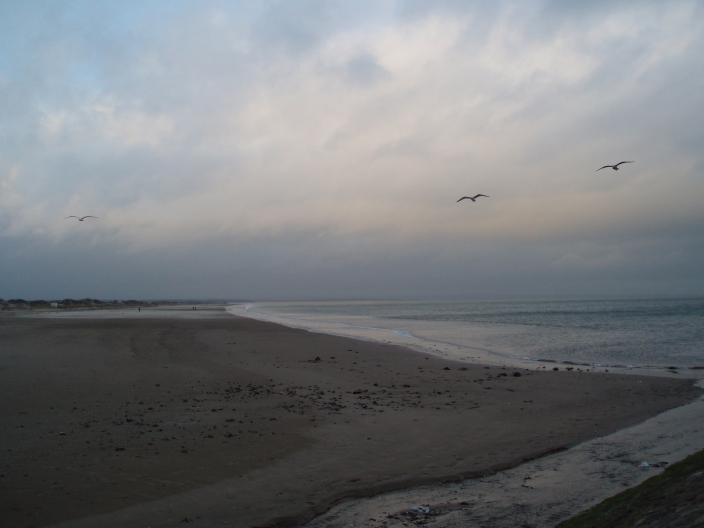} 
\end{figure}

\vfill

\footnotetext{The pictures of the title page and the conclusion are pictures of WestSands, a beach of St-Andrews (the town were I made my internship). They have no special meaning, they are just put here for their beauty.} 

\newpage

\pagestyle{plain}

~

\vspace{8cm}

\begin{center}

\section*{Acknowledgements}

\end{center}

I would like to thank my supervisor, Dr HongSheng Zhao, for his great attention in my work and more generally for having given me the opportunity of tackling with such interesting physics, his PhD students Xufen Wu and Garry W. Angus for interesting talks, my office mates: Rowan Smith for being really nice, Katharine Jonhstone for having taught me how to make green plants die as well, and Chris Poulton for his delightful way of talking. I would also like to thank Matthew J. S. Lee for being just great, and my family and friends far away for their support.

\newpage

\nocite{2}
\nocite{14}
\nocite{17}

\bibliographystyle{plain}
\bibliography{bibliorapsta}

\end{document}